\documentclass[prl,twocolumn,showpacs,preprintnumbers,amsmath,amssymb,superscriptaddress]{revtex4}

\usepackage{graphicx}
\usepackage{dcolumn}
\usepackage{bm}

\begin{document}

\title{Ferromagnetic $0-\pi$ Junctions as Classical Spins}

\author{M.L. Della Rocca}
\email{marilu@sa.infn.it}
\affiliation {Ecole Sup\`{e}rieure de
Physique et Chimie Industrielles (ESPCI), 10 rue Vauquelin, 75005
Paris, France}
\affiliation {Dipartimento di Fisica "E.R.
Caianiello", Universit\`{a} degli Studi di Salerno, via
S.Allende, 84081 Baronissi  (Salerno), Italy}
\author {M. Aprili}
\affiliation {Ecole Sup\`{e}rieure de Physique et Chimie
Industrielles (ESPCI), 10 rue Vauquelin, 75005 Paris, France}
\affiliation {CSNSM-CNRS, B\^{a}t.108 Universit\'{e} Paris-Sud,
91405 Orsay, France}
\author {T. Kontos}
\affiliation {Institute of Physics, University of Basel,
Klingelbergstrasse, 82, CH-4056 Basel, Switzerland}
\author {A. Gomez}
\affiliation {Ecole Sup\`{e}rieure de Physique et Chimie
Industrielles (ESPCI), 10 rue Vauquelin, 75005 Paris, France}
\author {P. Spatkis}
\affiliation {Ecole Sup\`{e}rieure de Physique et Chimie
Industrielles (ESPCI), 10 rue Vauquelin, 75005 Paris, France}

\date{\today}

\begin{abstract}
The ground state of highly damped PdNi based $0-\pi$ ferromagnetic Josephson
junctions shows a spontaneous half quantum vortex, sustained by a
supercurrent of undetermined sign. This supercurrent flows in the
electrode of a Josephson junction used as a detector and produces a
$\phi_{0}/4$ shift in
its magnetic diffraction pattern.
We have measured the statistics of the positive or
negative sign shift occurring at the superconducting transition of such a
junction. The randomness of the shift sign,
the reproducibility of its magnitude and the possibility of
achieving exact flux compensation upon field cooling: all these features
show that
$0-\pi$ junctions behave as classical spins, just as magnetic nanoparticles
with uniaxial
anisotropy.
\end{abstract}


\maketitle

Macroscopic devices formed by a large number of particles can
behave as a quantum two-level system provided phase coherence is
conserved at the macroscopic scale \cite{1,2}. One example is the
rf-SQUID, i.e. a superconducting ring interrupted by a Josephson
junction. In a rf$-$SQUID quantum tunneling between two different
macroscopic states corresponding to either zero or one quantum
flux in the ring has been observed in the past \cite{3,4}. More
recently, it was shown that when three Josephson junctions are
introduced in the ring, the ground state is two-fold degenerate
for an applied half quantum flux \cite{5}. These two macroscopic
configurations correspond to clockwise or anticlockwise
circulating supercurrent. Coherent superposition between these
two states has also been observed, making this device of great
interest for quantum electronics \cite{6,7,8}. However, as shown
below, clockwise or anticlockwise circulating supercurrent can
occur spontaneously (i.e. in zero applied magnetic field) when
the three junctions in the ring are replaced by one
$\pi-$junction ($\pi-$SQUID).

In this Letter, we show that the macroscopic ground state of a
ferromagnetic Josephson junction shorted by a 0 weak link mimics
that of a superconducting $\pi-$SQUID. Specifically, it is doubly
degenerate, so that half a quantum vortex $\phi_{0}/2$ or half a
quantum antivortex $-\phi_{0}/2$ appears in the junction. We
investigate the classical limit of this two-level system, which
behaves macroscopically as a magnetic nanoparticle of quantized
flux, the magnetic anisotropy axis being defined by the junction
plane.

$\pi-$junctions can be viewed as weak links characterized by an
intrinsic phase difference across them equal to $\pi$
\cite{9,10,11,12}. As a consequence, a $\pi-$junction in a
superconducting loop is a phase bias generator that produces a
spontaneous current and hence a magnetic flux \cite{13}. In the
limit $2\pi$$LI_{c}$$<$$\phi_{0}$, where $I_c$ is the junction
critical current and $L$ is the self-inductance of the loop, the
free energy is dominated by its magnetostatic part and its
minimum is reached for a zero total flux enclosed in the loop.
The system maintains a constant phase everywhere and a shift of
$\phi_{0}/2$ in the $I_{c}$($\phi$) relationship is expected
\cite{14,15}. When $2\pi$$LI_{c}$$>>$$\phi_{0}$, the Josephson
part of the free energy dominates and a phase gradient throughout
the loop favored thereby generating a spontaneous flux of $\pm
\phi_{0}/2$. The spontaneous supercurrent can circulate clockwise
and counterclockwise with exactly the same energy \cite{13}.
Applying a small magnetic field can lift this degeneracy and
defines an easy magnetization direction. The existence of a
spontaneous supercurrent sustaining half a quantum flux in
$\pi-$rings has been recently shown in Nb loops interrupted by a
ferromagnetic (PdNi) $\pi-$junction \cite{16}. Direct scanning
SQUID microscope imaging of the half-integer vortex has also been
reported in HTSC grain boundary junctions \cite{17,18}.
Analogously, a highly damped single Josephson junction fabricated
with a $0$ and a $\pi$ region in parallel should display a
spontaneous half quantum vortex at the $0-\pi$ boundary
\cite{13}.
\begin{figure}
  \includegraphics[width=0.45\textwidth]{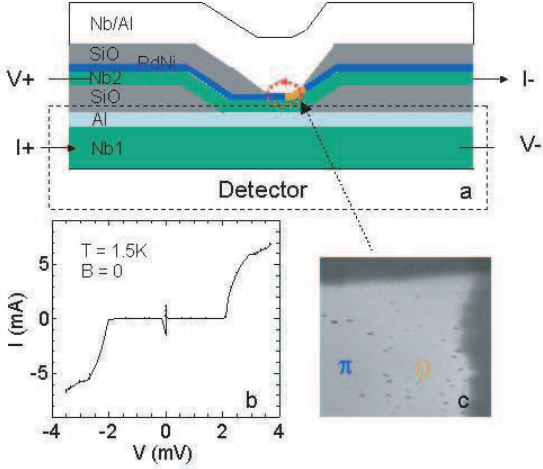}\\
  \caption{(a) Schematic cross section of the device: a Nb based Josephson
junction (detector) is coupled to a ferromagnetic (PdNi) junction
(source) by sharing one electrode (Nb2). The red-colored closed
loop indicates the half quantum vortex's location. (b) I-V
characteristic of the detector at 1.5K in zero applied external
field. (c) Scanning electron microscope image of the
ferromagnetic junction edge area. The black spots are
inhomogeneities which prevent PdNi continuity inducing a $0$-
coupling region.}
\end{figure}

We detect a spontaneous supercurrent by measuring the phase
gradient with another Josephson junction. The ferromagnetic
junction (source) and the detection junction (detector) are
coupled, as shown in Fig.1(a), by sharing an electrode. I.e., the
top electrode of the conventional Josephson junction is
simultaneously the bottom electrode of the ferromagnetic one. If
half a quantum vortex is spontaneously generated in the
ferromagnetic junction, the spontaneous supercurrent that
sustains it circulates in the common electrode [Nb2, Fig.1(a)]
producing a phase variation equal to $\pi/2$. A
$\phi_{0}/4-$shift of the detection junction's diffraction
pattern is thus produced. When an external magnetic field is
applied, the diffraction pattern of the detection junction is
given by:
\begin{equation}
  I(B)=I(0)\frac{sin\left[\frac{\pi}{\phi_{0}}(k_{s}J_{s}+k'\phi'+\phi)\right]}
  {\left[\frac{\pi}{\phi_{0}}(k_{s}J_{s}+k'\phi'+\phi)\right]}
\end{equation}
where $\phi'=BDt'$ and $\phi=BDt$ are the magnetic fluxes through
the ferromagnetic and detection junction respectively, with $D$
the junction width, $t$ and $t'$ the effective barrier thickness.
$J_{s}$ is the spontaneous supercurrent density,
$k_{s}=(\frac{\mu_0\lambda_L(T)^2}{2})D$ and
$k'=(\frac{\mu_0\lambda_L(T)^2}{L})\frac{D}{wd_{Nb2}}$, with
$\mu_{0}$ the vacuum permittivity, $\lambda_{L}(T)$ the Nb London
penetration depth, $L$ the ferromagnetic junction inductance, $w$
the junction length and $d_{Nb2}$ the thickness of the common
electrode, assumed to be of the order of the London penetration
depth. The term $k_{s}J_{s}$ generates the shift due to the
spontaneous supercurrent contribution, while the term $k'\phi'$
reduces the diffraction pattern period as a result of the
contribution due to the screening current in the ferromagnetic
junction.

Samples are fabricated by e-gun evaporation in an ultra high
vacuum (UHV) system in a typical base pressure of $10^{-9}$mbar.
The whole device is fabricated completely in situ by shadow
masks. The maximum degree of misalignment is $100\mu$m.
Deposition rates are monitored by a quartz balance with
$0.1{\AA}$/s resolution. First the bottom planar
Nb/Al/Al$_2$O$_3$/Nb detection junction is made. A $1000{\AA}$
thick Nb strip [Nb1, Fig.1(a))] is evaporated and backed by
$500{\AA}$ of Al. An Al$_2$O$_3$ oxide layer is achieved by
oxygen plasma oxydation during $12$ min, completed in a $10$ mbar
O$_{2}$ partial pressure during $10$ min. A square window of
$0.6\times0.8$mm$^2$ ($D\times w$), obtained by evaporating
$500{\AA}$ thick SiO layers, defines the junction area. Then, a
$500{\AA}$ thick Nb layer [Nb2, Fig.1(a)] is evaporated
perpendicular to the Nb/Al strip to close the junction. This
procedure results in a junction critical temperature, $T_{cj}$,
equal to $8.5$K. Typical junction normal state resistances are of
the order of $0.1-1\Omega$ and critical current values are of
$1-10$mA at $4.2$K. The resulting critical current density is
$\sim1-10^{-1}$ A/cm$^{2}$ leading to a Josephson penetration
depth $\lambda_{j}\geq1$mm, i.e. larger than the size of the
junction (small limit). The I-V characteristic of a typical
detector is shown in Fig.1(b). The Nb2 layer acts as both the
counterelectrode of the bottom detection junction and the base
electrode of the top ferromagnetic junction. Its thickness is
comparable to the Nb penetration depth to insure good coupling
between the two junctions. The same procedure is used to prepare
the top planar Nb/PdNi/Nb/Al junction. Specifically, after
defining the same junction area by evaporating $500{\AA}$ thick
SiO layers, a PdNi layer was evaporated directly on the Nb layer,
without any Al-oxide barrier. This results in a very large
critical current and very small junction resistance \cite{19}. An
estimate of the critical current density is $10^{4}-10{^5}$
A/cm$^{2}$, so the Josephson penetration depth
$\lambda_{jf}<10^{-2}$mm$\ll$$D$. Hence the "source"
ferromagnetic junction is in the "large limit" with a large
screening capability. The junction was closed by a $100{\AA}$
thick Nb layer backed by a $750{\AA}$ thick Al layer, resulting
in a critical temperature, $T_{cf}$, of $3.6$K. As the critical
temperature of the detector was higher than that of the source,
its diffraction pattern could be measured with and without
spontaneous supercurrents. The Ni concentration was checked by
Rutherford backscattering spectrometry on PdNi samples evaporated
in the same run as the junctions and was equal to $9\%$,
corresponding to an exchange energy, E$_{ex}$, of $10.5$meV and a
$0$ to $\pi-$coupling transition for d$_{0-\pi}\sim75{\AA}$.

An intrinsic feature of our fabrication technique is the
formation of inhomogeneities at the window edges of the
ferromagnetic junction [see Fig.1(c)]. SEM images and AFM
analysis \cite{20} have revealed bubbles [black spots in
Fig.1(c)] with diameter of about $2-3{\mu}m$ and roughness up to
$1000{\AA}$ much higher than the PdNi thickness. These
inhomogeneities have been observed in all samples and they may
produce shorts through the ferromagnetic layer resulting in a
$0-$coupling at the junction edge, independently from the PdNi
layer thickness. Therefore, $0-$junctions and $0-\pi$ junctions
are obtained with PdNi layer thickness corresponding to
$0-$coupling and $\pi-$coupling, respectively.
\begin{figure}
  \includegraphics[width=0.35\textwidth]{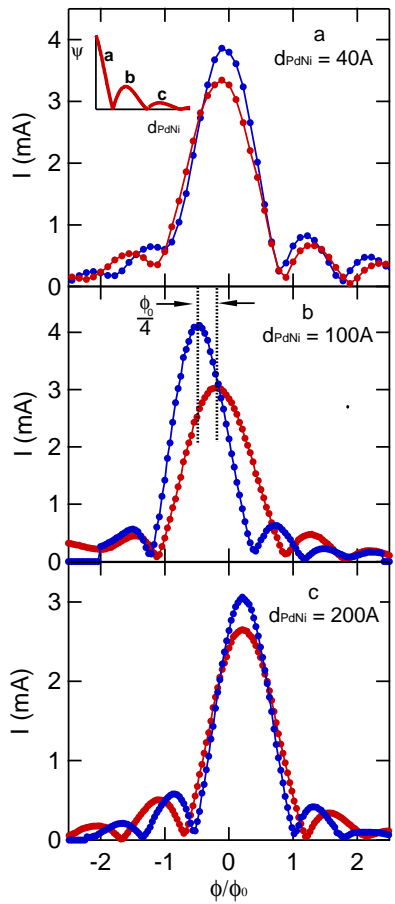}\\
  \caption{Detector diffraction patterns for sources
with d$_{PdNi}\sim40{\AA}-100{\AA}-200{\AA}$, corresponding
respectively to $0-$coupling (a)$-$(c) and $\pi-$coupling (b).
Measurements are taken at $T=4.2$K$>$$T_{cf}$ (red data) and
$T=2$K$<$$T_{cf}$ (bleu data). A $\phi_{0}/4-$shift of the
maximum critical current appears in the (b) case, where a $0-\pi$
coupling region is realized at the ferromagnetic junction edge.
Inset: Oscillating behaviour of the order parameter as function
of d$_{PdNi}$.}
\end{figure}

Measurements were made in a $^4$He flow cryostat. Residual fields
were screened by cryoperm and $\mu$-metal shields. All
measurements showed comparable residual fields at $4.2$K of some
tenths of mG. Depending on the PdNi layer thickness, the
ferromagnetic junction is either $0$ or $\pi$ while a spontaneous
half quantum vortex or antivortex is expected only for
$\pi-$coupling. This is the main result of our experiment as
reported in Fig.2, where we show the diffraction patterns of
three samples for PdNi thickness equal respectively to $40{\AA}$,
$100{\AA}$ and $200{\AA}$ at $T=4.2$K (red data) and $T=2$K (bleu
data). For $0-$coupling [$40{\AA}$ and $200{\AA}$, Fig.2(a) and
Fig.2(c) respectively] at $T<T_{cf}$, the period is reduced but
no shift occurs in the detector diffraction pattern. The magnetic
field corresponding to a quantum flux is $300$mG. On the other
hand, for $\pi-$coupling [$100{\AA}$, Fig.2(b)] the period
reduction is accompanied by a shift of $\phi_{0}/4$ in the
detector, as expected for a spontaneous half quantum vortex in
the ferromagnetic junction. The period reduction at the lowest
temperature ($2$K) for all the samples was about $15\%$. This
value results from two competing effects: a smaller period is
expected because of screening in the ferromagnetic junction as
shown in eq.1, whereas the decrease in the penetration depth,
$\lambda(T)$, at lower temperatures should increase the
modulation period. We found no period reduction or diffraction
pattern shift when the source and the detector were decoupled by
a thin insulating layer. This indicates that ordinary inductive
coupling between the two junctions is negligible. Regarding the
possible effect of an external residual field or the magnetic
layer itself on the spontaneous supercurrents, it is important to
stress that the shifts are always about $\phi_{0}/4$ for
$\pi-$coupling and zero for $0-$coupling. Since this depends
neither on the residual field nor, for $0-$coupling, on the
thickness of the ferromagnetic layer, any effect of the PdNi
magnetic moment on the amplitude of the spontaneous supercurrents
can be ruled out. The PdNi magnetic structure only lifts the
degeneracy of the ground state and polarizes the supercurrents.
As a consequence, the sign shift is always the same below the
Curie temperature ($\sim100$K), indicating the same spontaneous
current polarization.
\begin{figure}
  \includegraphics[width=0.35\textwidth]{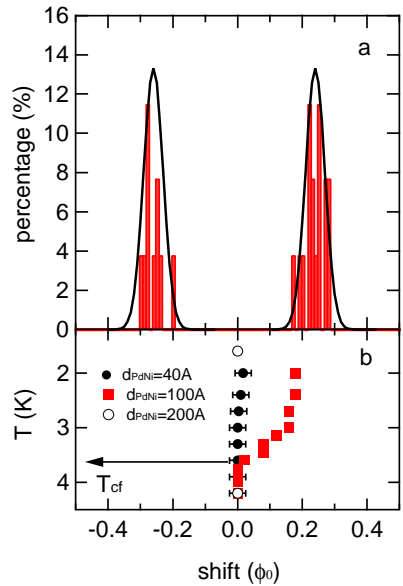}\\
  \caption{(a) Amplitude histograms (red bars) of the spontaneous shift
for 26 different cooldowns from room temperature of different
samples, and their best Gaussian fits (black curves). The
Gaussians have mean values of $+0.24\phi_{0}$ and
$-0.26\phi_{0}$, with dispersions equal to $\pm0.03\phi_{0}$. (b)
Temperature dependence of the spontaneous shift for the samples
of Fig.2: d$_{PdNi}\simeq40{\AA}$ (black circles),
d$_{PdNi}\simeq100{\AA}$ (red square) and
d$_{PdNi}\simeq200{\AA}$ (open circles). The spontaneous half
quantum flux appears below the ferromagnetic junction
superconducting transition temperature, $T_{cf}$.}
\end{figure}

When cooling down from room temperature to $2$K, the shift, while
reproducible in magnitude, becomes random in sign as shown in
Fig.3(a). The gaussian distribution functions used to approximate
these distributions show mean values of $+0.24\phi_{0}$ and
$-0.26\phi_{0}$ with equal dispersions of $\pm0.03\phi_{0}$. This
is the expected behavior of a two$-$fold degenerate ground state
corresponding to either half a quantum flux or half a quantum
antiflux in the ferromagnetic junction. The same distribution
would be expected for a magnetic nanoparticle with a significant
uniaxial magnetic anisotropy. The large dispersion in shifts can
be related to an intrinsic limit of the device itself since the
same dispersion is observed when measuring devices with uniform
$0-$coupling ferromagnetic junction, where the mean shift value
is $0$. At 2K, the thermal activation energy ($K_{b}T\simeq
0.2$meV) is negligible with respect to the Josephson energy
($E_{J}=\frac{I_{c}\phi_{0}}{2\pi}\simeq10^{5}-10^{6}$eV for
$J_{c}\sim10^{4}-10^{5}$A/cm$^{2}$ and a junction area, $D\times
w$, of $0.48$mm$^{2}$) and hence no hopping between the two
potential walls can occur. Fig.3(b) shows the temperature
dependence of the shift measured for the junctions whose results
are presented in Fig.2. A finite shift is observed starting from
$3.6$K ($T_{cf}$). The shift saturates at low temperature, just
as the critical current does. Fig.3(b) also shows that the shift
remains zero, as expected, for $0-$junctions.
\begin{figure}
  \includegraphics[width=0.4\textwidth]{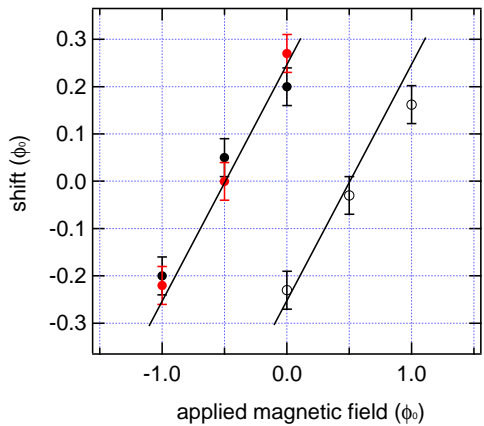}\\
  \caption{Field cooled measurements for a half and
an integer quantum flux applied to the ferromagnetic junction in
the direction opposite to the spontaneous half quantum flux
during cooldown through $T_{cf}$.}
\end{figure}

Finally we studied the polarization of the spontaneous
supercurrents when an external magnetic field is applied during
cooling down through $T_{cf}$ (Field-Cool (FC)). This field
breaks time reversal symmetry and lifts the ground state
degeneracy. In Fig.4 we show the diffraction pattern shift as a
function of the applied flux for either a half- or an integer
quantum flux in the ferromagnetic junction. For zero-field
cooling (ZFC), the shift is either $+\phi_{0}/4$ (red and black
circles) or $-\phi_{0}/4$ (open circles). When cooling occurs
with a $\phi_{0}/2$ flux in the direction opposite to the
spontaneous flux, screening currents in the ferromagnetic
junction are induced, which compensate exactly for the half
quantum vortex, so that no shift results. When cooling with a
flux $\phi_{0}$ in the direction opposite to the spontaneous
flux, the screening current and the spontaneous ones add up,
inducing a net flux equal to half a quantum in the direction
opposite to the spontaneous one. The two lines, in Fig.4, define
the slip-over from  vortex to antivortex and vice-versa under FC.

In conclusion, we have probed the ground state of highly damped
ferromagnetic $0-\pi$ junctions by a phase sensitive technique.
The samples were directly coupled to a detection junction with a
higher critical temperature via a shared electrode. The
spontaneous supercurrent, sustaining half a quantum vortex in the
$0-\pi$ junction, produced a $\phi_{0}/4-$shift in the detection
junction diffraction pattern below the ferromagnetic junction
transition. Although equal in magnitude, the shifts were random
in direction for different cooldowns, as a result of the doubly
degenerate ground state of the $0-\pi$ junction, corresponding to
equal vortex or antivortex probabilities. Thus a $0-\pi$
Josephson junction is a macroscopic realization of a two-level
system and behaves in the classical limit as a single magnetic
domain with uniaxial anisotropy. It would be interesting to
investigate the quantum limit by reducing the temperature and the
barrier height between the two potential wells.

We are indebted to A. Bauer and C. Strunk for suggesting the
field-cooled compensation and E. Reinwald for performing a detailed AFM
study on
our samples. We also thank H. Pothier, A. Buzdin, E. Goldowin, J.
Lesueur for stimulating discussions; S. Collin for technical
assistance and H. Bernas for a critical reading of the manuscript.
This work has been partially founded by ESF through the
"$\pi-$shift" program.

\end{document}